\documentstyle[12pt]{article}

\newcommand\mysection{\setcounter{equation}{0}\section}

\newskip\humongous \humongous=0pt plus 1000pt minus 
1000pt

\newif\ifdtup

\def\figcap{\section*{Figure Captions\markboth
        {FIGURECAPTIONS}{FIGURECAPTIONS}}\list
        {Figure 
\arabic{enumi}:\hfill}{\settowidth\labelwidth{Figure
999:}
        \leftmargin\labelwidth
        \advance\leftmargin\labelsep\usecounter{enumi}}}
 \relax

\def\bom#1{{\mbox{\boldmath $#1$}}}
\def\beq#1{\begin{equation}\label{#1}}
\def\beeq#1{\begin{eqnarray}\label{#1}}

\def\eeq{\end{equation}}
\def\eeeq{\end{eqnarray}}
\def\ltap{\raisebox{-.4ex}{\rlap{$\,\sim\,$}} \raisebox{.4ex}{$\,<\,$}}

\def\as{\alpha_S}
\def\x{x_{\gamma}}
\def\r{r_{\gamma}}
\def\d{\delta}
\def\asm{\frac{\alpha_S(\mu^2)}{2\pi}}

\def\vep{\varepsilon_h}

\def\s#1{{\small #1}}

\def\LO{\s{LO}}
\def\NLO{\s{NLO}}
\def\msbar{$\overline {\rm {\s MS}}$}

\def\irlim{\raisebox{-1.2ex}{\rlap{\tiny $\;\;\;p_i \to 0$}}
\raisebox{0ex}{$\;\;\;\,\to\;\;\;\,$}}

\def\colim{\raisebox{-1.2ex}{\rlap{\tiny $\;\;\;p_i \parallel p_j$}}
\raisebox{0ex}{$\;\;\;\,\to\;\;\;\,$}}

\def\cflim{\raisebox{-1.2ex}{\rlap{\tiny $\;\;\;p_i \parallel p_\gamma$}}
\raisebox{0ex}{$\;\;\;\,\to\;\;\;\,$}}

\def\cflimn{\raisebox{-1.2ex}{\rlap{\tiny $\;p_{n+1} \parallel p_\gamma$}}
\raisebox{0ex}{$\;\;\;\,\to\;\;\;\,$}}

\def\np#1#2#3{Nucl.\ Phys.\ B#1 (19#3) #2}
\def\pl#1#2#3{Phys.\ Lett.\ #1B (19#3) #2}
\def\pr#1#2#3{Phys.\ Rev.\ D #1 (19#3) #2}

\def\prl#1#2#3{Phys.\ Rev.\ Lett.\ #1 (19#3) #2}

\def\zp#1#2#3{Z.\ Phys.\ C#1 (19#3) #2}

\def\ej#1#2#3{JHEP\ #1 (19#3) #2}

\begin{document}

\begin{titlepage}
\renewcommand{\thefootnote}{\fnsymbol{footnote}}
\begin{flushright}
     CERN--TH/98-82 \\ LPTHE--ORSAY 98/10 \\ LAPTH--675/98
     \\ hep-ph/9803475  
%     \\ \today
     \end{flushright}
\par \vspace{10mm}
\begin{center}
{\Large \bf Factorization and Soft-Gluon Divergences in \\[1ex]
Isolated-Photon Cross Sections\footnote{This work was supported in part 
by the EU Fourth Framework Programme ``Training and Mobility of Researchers'', 
Network ``Quantum Chromodynamics and the Deep Structure of
Elementary Particles'', contract FMRX--CT98--0194 (DG 12 -- MIHT).}}
\end{center}
\par \vspace{2mm}
\begin{center}
     {\bf S. Catani}~\footnote{On leave of absence from INFN,
     Sezione di Firenze, Florence, Italy.}\\
     {Theory Division, CERN, CH 1211 Geneva 23, Switzerland} 
%INFN, Sezione di Firenze}\\ {
%and Dipartimento
%di Fisica, Universit\`a di Firenze}\\
%{Largo E. Fermi 2, I-50125 Florence, Italy}\\
%     \par \noindent
%     and
     \par \noindent
     {\bf M.\ Fontannaz} \\
     LPTHE, Universit\'{e} Paris-Sud, B\^{a}timent 211, 
     F-91405 Orsay Cedex, France
     \par \noindent
     and
     \par \noindent
     {\bf E.\ Pilon}~\footnote{On leave of absence from LPTHE, 
     Universit\'{e} Paris-Sud, Orsay Cedex, France.}
     \\
     LAPTH, B.P. 110, F-74941 Annecy-le-Vieux Cedex, France 
\end{center}
\par \vskip 2mm

\begin{center} {\large \bf Abstract} \end{center}
\begin{quote}
\pretolerance 10000

We study the production of isolated photons in $e^+e^-$ annihilation and
give the proof of the all-order factorization of the collinear
singularities. These singularities are absorbed in the standard 
fragmentation functions of partons into a photon, while
the effects of the isolation are
consistently included in the short-distance cross section. We compute this
cross section at order $\as$
and show that it contains 
large double logarithms of the isolation parameters. 
We explain the physical origin
of these logarithms and discuss the possibility to resum them to all
orders in $\as$.

\end{quote}

\vspace*{\fill}
\begin{flushleft}
     CERN--TH/98-82 \\  LPTHE--ORSAY 98/10 \\ LAPTH--675/98
     \\   March 1998
\end{flushleft}
\end{titlepage}

\renewcommand{\thefootnote}{\fnsymbol{footnote}}

\mysection{Introduction}
\label{intro}

At LEP, the Tevatron and the future LHC, the detection of high-energy photons 
produced by short-distance interactions
goes through the definition of an isolation criterion that aims at decreasing 
fake photon signals coming from $\pi^0$ decays. 
The high-energy $\pi^0$, which is 
seen as a single cluster in a calorimeter, belongs to a jet and is
accompanied by hadronic energy, whereas a bremsstrahlung photon emitted by 
a quark is unaccompanied as soon as the photon--quark angle is large 
enough. Therefore any criterion to select `unaccompanied photons'  
enhances the signal--background ratio. 

The criterion that is mostly used in current practice is the following.
In $e^+e^-$ collisions at LEP,
a photon of energy $E_{\gamma}$ is said to be isolated if it is accompanied by 
less 
than a specific amount $\vep E_{\gamma}$ of hadronic energy $E_h^{\rm cone}$
in a cone of half-angle $\delta$ around the photon momentum. This
isolation criterion can be written as follows:
\beq{pic}
E_h^{\rm cone} = \sum_{i\not= \gamma} E_i \;\Theta(\delta - \theta_{i\gamma}) 
\leq \varepsilon_h E_{\gamma} \;,  \;\;\;\; \vep > 0 \;,
\eeq
where the energies and the relative angles $\theta_{i\gamma}$ are defined in 
the centre-of-mass frame. In hadron collisions at the Tevatron, the isolation
criterion is similar, but the energies are replaced by the transverse
energies 
%with respect to the beam direction 
while the angles and the cone are defined
in the azimuth--pseudorapidity plane, so that $\theta_{i\gamma}$ is
replaced by $R_{i\gamma}= \sqrt {\eta^2_{i\gamma} + \phi^2_{i\gamma}}$.
Typical values of the isolation parameters $\vep$ and $\d$ are
$\vep = {\cal O}(10^{-1})$ and $\delta \ltap 0.7$.

The criterion in Eq.~(\ref{pic}) can be stated in an
equivalent way by saying that the fraction of electromagnetic energy
inside the isolation cone has 
to be larger than a fixed value $x_c$:
\beq{pixc}
\frac{E_\gamma}{\sum_{i\neq \gamma} E_i \; \Theta(\d - \theta_{i\gamma})
+ E_\gamma} \geq x_c \;\;,
\eeq
where
\beq{xc}
x_c \equiv \frac{1}{1+\epsilon_h} < 1 \;\;.
\eeq

In the case when no isolation criterion is applied, the inclusive photon 
cross section is computable by using the QCD factorization formula:
\beq{qcdfac}
\sigma_{\gamma} = \sum_{A=q, \bar{q}, g, \gamma} C_A (\mu^2) 
\otimes D_{\gamma/A}(\mu^2 ) \;, 
\eeq 
where $\otimes$ denotes the convolution over the energy fraction $z$. 
The hard-subprocess cross section $C_A(\mu^2;z)$,
which des\-cri\-bes the production of a high-energy 
parton $A$, can be calculated order by order in QCD perturbation theory.
The non-perturbative phenomena are included in $D_{\gamma/a}(z,\mu^2)$, 
the inclusive fragmentation function of the QCD parton 
$a$ ($a=q, \bar{q}, g$) into a photon \cite{bourhis, grvphot}, and the
direct term ($A=\gamma$) not proportional to the 
fragmentation function is taken into account through the convention 
$D_{\gamma/\gamma}(z, \mu^2) = \delta (1 - z)$.

In the case of an isolated photon, the isolation criterion 
(\ref{pic}) enforces additional phase-space restrictions. This implies that 
the cross section is {\em no} longer fully inclusive and, hence, that
the factorized expression (\ref{qcdfac}) is {\em not} necessarily valid.

The applicability of the factorization theorem to isolated photons is thus
a basic issue that has to be dealt with before the corresponding
cross sections can be studied within the conventional QCD framework.
The perturbative calculations that have been performed so far do not help
to this purpose. In the case of hadron collisions, all the available
calculations \cite{hadcoll} beyond the leading order (\LO)
in the strong coupling $\as$ are 
based on approximate methods that, in particular,
lead to an incomplete treatment of the fragmentation contributions.
The next-to-leading order (\NLO) calculation of Ref.~\cite{zoltan} 
for $e^+e^-$ collisions introduces by
definition a fragmentation component that explicitly depends on the isolation
parameters and thus differs from the process-independent fragmentation
function in Eq.~(\ref{qcdfac}). Actually, in the case of 
$e^+e^-$ annihilation, factorization has recently been questioned by
Berger, Guo and Qiu \cite{berger}. The results of Ref.~\cite{berger}
have been criticized by some of us in Ref.~\cite{AFGKP}.

In the present paper we confirm that factorization is fulfilled
at \NLO\ \cite{AFGKP} and extend this argument 
by proving the validity of factorization to all orders in
perturbation theory.

Although we show that isolation does not spoil
factorization, yet the phase-space 
restrictions due to the isolation criterion (\ref{pic}) are not harmless,
as discussed in detail in the second part of the paper.
The factorization theorem deals with collinear singularities that occur
in the calculation at the parton level. Once these singularities, whose origin
is non-perturbative, have been absorbed in the fragmentation functions,
the short-distance cross section can still have a divergent behaviour at 
{\em some points} of the phase space when computed order by order 
in perturbation theory \cite{CW}. These divergences are due to certain
kinematical constraints that, limiting the fully-inclusive character of the 
cross section, produce an imperfect compensation between real and virtual
emission of soft (and collinear) partons. In isolated-photon cross sections,
the soft-gluon divergences are double logarithmic and appear at a specific
point {\em inside} the phase space \cite{berger, AFGKP}. 
Owing to its perturbative origin, this disease can be cured by summing 
the logarithmic divergences to all orders in perturbation theory. 

The outline of the paper is as follows. In Sect.~\ref{factsec} we give our
proof of the validity of the factorization theorem for isolated-photon 
cross sections defined by the criterion\footnote{The case of jets containing
isolated photons has been considered in 
Refs.~\cite{zoltan, glover, phpjet}. An alternative
definition of the isolated photon has recently been suggested by 
Frixione \cite{frixione}.} in Eq.~(\ref{pic}). We show how 
the effects of the isolation are consistently included in the
short-distance subprocess and, in particular, we discuss the
functional dependence of the short-distance cross section on the isolation
parameters. In Sect.~\ref{fixor} we consider isolated photons produced in
$e^+e^-$ annihilation and we compute the \NLO\ contribution to the
fragmentation component of the short-distance cross section. We present
results in analytic form for any value of the isolation parameters $\vep, \d$.
Using these explicit expressions, in Sect.~\ref{divsec} we 
discuss in detail the physical origin of
the soft-gluon divergent behaviour of the \NLO\ cross section in the vicinity
of point $x_\gamma = x_c$. Finally, in Sect.~\ref{finsec} we summarize our
results and outline how an all-order resummation can eventually lead to
well-behaved theoretical predictions for the cross section.

\mysection{Factorization}
\label{factsec}

The factorization issue regards both hadron and photon
distributions. Thus, we make no distinction between these two cases,
although isolated-hadron cross sections are of less experimental interest.

\subsection{$e^+e^-$ annihilation}
\label{eesect}

The customary factorization formula for the inclusive
distribution of a single particle $H$ with four-momentum $p_\gamma$ produced in
$e^+e^-$ annihilation is:
\beq{facforfull}
\!\!\!\!\!\!
\frac{1}{\sigma_0} 
\frac{d\sigma(Q^2, \x)}{d\x} =
\sum_{A} \; \int_{\x}^1 \frac{dx}{x} \;
\; D_{H/A}(\x/x,\mu^2) \;
C_A^{({\rm full})}(\as(\mu^2),Q^2/\mu^2;x) + {\cal O}\left((1/Q)^p\right) \;,
\eeq 
where the sum extends over $A=q_f,{\bar q}_f,g$ when the observed particle 
$H$ is a hadron and over $A=q_f,{\bar q}_f,g,\gamma$ when $H=\gamma$ is a photon
(in this case $D_{\gamma/\gamma}(z,\mu^2) = \delta(1-z)$,
by definition). The term ${\cal O}\left((1/Q)^p\right)$ on the right-hand
side denotes corrections that are suppressed by some power $p \geq 1$ of
the centre-of-mass energy $Q$ when $Q \gg \Lambda_{QCD}$.

The formula (\ref{facforfull}) states that at large values of $Q^2$
and for any fixed value of the energy fraction $\x=2p_\gamma \cdot Q/Q^2$, 
all the long-distance physics phenomena can be absorbed in the non-perturbative
fragmentation function $D_{H/A}(z,\mu^2)$ of the parton $A$ into the particle
$H$. The remaining coefficient function 
$C_A^{({\rm full})}(\as(\mu^2),Q^2/\mu^2;x)$ is short-distance-dominated 
and depends only on the partonic subprocess. 

The predictivity of Eq.~(\ref{facforfull}) within perturbation theory 
follows from the fact that not only $C_A^{({\rm full})}$ but 
also the $Q^2$-evolution of the fragmentation functions 
are perturbatively computable as power-series expansions in $\as$. 
After having extracted $D_{H/A}(z,Q_0^2)$
from experimental data at a certain value $Q^2=Q^2_0$, perturbative QCD
predicts the cross section in Eq.~(\ref{facforfull}) for any other value of 
$Q^2$.

The content of the factorization theorem in perturbation theory is
nonetheless wider. It states that similar factorization formulae are
valid for other {\em less inclusive} observables and 
that, in these formulae, the 
dominant non-perturbative contribution is {\em universal} and accounted for
by the same fragmentation functions $D_{H/A}(z,\mu^2)$ as in 
Eq.~(\ref{facforfull}). Thus, the factorization formulae are obtained from
(\ref{facforfull}) by the replacement
\beq{fulltoJ}
C_A^{({\rm full})}(\as(\mu^2),Q^2/\mu^2;x) \to
C_A(\as(\mu^2),Q^2/\mu^2;x, \{J\}) \;,
\eeq
where $\{J\}$ denotes the dependence on the particular observable.
The differences among the various observables only regard the perturbatively
computable coefficient functions $C_A$ and the size and the power $p$ of
the power-suppressed corrections.

A particular class of observables to which the factorization theorem applies
is the class formed by what we call jet-type observables. These 
observables can be easily identified by examining how they
are defined (and measured) in terms of the momenta of the final-state 
particles in the process.
The definition has to fulfil the requirements
of $i)$~infrared safety, $ii)$~collinear safety, and $iii)$~collinear
factorizability. 

The first two requirements
regard the dependence on all particle momenta
but the triggered momentum $p_\gamma$: infrared safety
means that the value of the observable is independent of the momenta of 
arbitrarily soft particles, and collinear safety
implies that, when some final-state
particles are produced collinearly, the value of the observable
depends on their total momentum rather than on the momentum of each of
them. Collinear factorizability means that from the measurement
of the observable one cannot distinguish whether  
the triggered momentum $p_\gamma$ is carried by the particle $H$ or by
that particle accompanied by a bunch of particles parallel to it.

These properties can be stated in a formal way as follows \cite{submeth, slice,
dipmeth}. Let us first 
introduce the exclusive cross section 
$d\sigma_n^{(\rm excl.)}(p_\gamma,p_1,\dots,p_n)$ to produce $n+1$ particles
with momenta $p_\gamma,p_1,\dots,p_n$, so that the single-particle inclusive 
distribution in Eq.~(\ref{facforfull}) can be written in the following form
\beq{fullxs}
\!\!\!\!\!\!
\frac{d\sigma(Q^2, \x)}{d\x} = \sum_n \int_{\Omega(p_\gamma,p_1,\dots,p_n)}
d\sigma_n^{(\rm excl.)}(p_\gamma,p_1,\dots,p_n) 
\;\delta(\x - 2 p_\gamma \cdot Q/Q^2) \;,
\eeq 
where the integration extends over the full $(n+1)$-particle phase space
$\Omega(p_\gamma,p_1,\dots,p_n)$. According to this notation, any
less inclusive cross section $d\sigma_J$ is given by
\beeq{Jxs}
\!\!\!\!\!\!
\frac{d\sigma_J(Q^2, \x)}{d\x} = \!\!\!&&\!\!\!
\sum_n \int_{\Omega(p_\gamma,p_1,\dots,p_n)}
d\sigma_n^{(\rm excl.)}(p_\gamma,p_1,\dots,p_n) \nonumber \\
&& \;\; \cdot \;\delta(\x - 2 p_\gamma \cdot Q/Q^2) \;
F_J^{(n)}(Q,p_\gamma,\{J\};p_1, \dots p_n) \;,
\eeeq 
where, for any
exclusive final state with $n+1$ particles of momenta 
$p_\gamma, p_1, \dots p_n$,
we have denoted by
\beq{fn} 
F_J^{(n)}(Q,p_\gamma,\{J\};p_1, \dots p_n)
\eeq 
the measurement function that defines the actual observable.

In terms of this function, the properties of a jet-type observable 
are\footnote{In Eqs.~(\ref{irsafe})--(\ref{cfact}) bold-face characters are used
just to emphasize the differences between left-hand and right-hand sides.}

\noindent $i)$ infrared safety:
\beq{irsafe}
F_J^{(n{\bom {+1}})}(Q,p_\gamma,\{J\};p_1, \dots {\bom p_i}, \dots p_{n+1}) 
%\to 
\irlim
F_J^{(n)}(Q,p_\gamma,\{J\};p_1, \dots p_{n+1}) \;,
\eeq

\noindent $ii)$ collinear safety:
\beq{csafe}
F_J^{(n{\bom {+1}})}(Q,p_\gamma,\{J\};p_1, \dots {\bom p_i}, 
{\bom p_j}, \dots p_{n+1}) 
\colim 
F_J^{(n)}(Q,p_\gamma,\{J\};p_1, \dots, {\bom {p_i + p_j}}, \dots p_{n+1}) \;,
\eeq

\noindent $iii)$ collinear factorizability:
\beq{cfact}
F_J^{(n{\bom {+1}})}(Q,{\bom p_\gamma},\{J\};p_1, \dots {\bom p_i}, \dots 
p_{n+1}) 
\cflim
F_J^{(n)}(Q,{\bom {p_\gamma+p_i}},\{J\};p_1, \dots \dots p_{n+1}) \;.
\eeq

In the case of the isolated-particle distribution, the measurement function
is specified by the isolation criterion in Eq.~(\ref{pic}). One
sees that the expression (\ref{pic}) fulfils
the properties (\ref{irsafe}) and (\ref{csafe}), provided the isolation
parameter $\varepsilon_h$ is kept finite\footnote{The violation of infrared
safety in the case of perfect isolation, i.e. when $E_h^{\rm cone}=0$,
was pointed out in Refs.~\cite{zoltan, glover, bergerqiu}.}.

The fulfilment of the collinear factorizability in Eq.~(\ref{cfact})
may appear more problematic by naive inspection of Eq.~(\ref{pic}). 
If the isolation parameters $J = \{ \varepsilon_h, \delta \}$ were assumed to
be the relevant variables for the short-distance subprocess, 
the measurement function could be defined as 
\beq{fwrong}
F^{(n+1)}(Q,p_\gamma,J=\{ \varepsilon_h, \delta \};p_1, \dots p_{n+1}) = 
\Theta\!\left( \vep E_{\gamma} - 
\sum_{\stackrel{i=1}{i\neq \gamma}}^{n+1} E_i \; \Theta(\d -
\theta_{i\gamma}) \right) \;\;.
\eeq
Then, considering the collinear limit where, for instance, $p_{n+1}$ is
parallel to $p_\gamma$: 
\beeq{fwronglim'}
\!\!\!\!\!\!\!\!\!\!\!\!
F^{(n+1)}(Q,p_\gamma,J=\{ \varepsilon_h, \delta \};p_1, \dots p_{n+1}) \!\!\!
\!\!\!\! \!\!\!&\!\!&\cflimn 
\Theta\!\left( \vep E_{\gamma} - E_{n+1} - 
\sum_{\stackrel{i=1}{i\neq \gamma}}^{n} E_i \; \Theta(\d -
\theta_{i\gamma}) \right) \nonumber \\
\label{fwronglim}
\!\!\!\!\!\!\!&\!\!\!\!\!& = 
F^{(n)}(Q,p_\gamma- p_{n+1}/\vep,J=\{ \varepsilon_h, \delta \};p_1,\dots p_{n})
\,.
\eeeq
Since the function $F^{(n)}$ on the right-hand side of Eq.~(\ref{fwronglim})
depends on the momentum $p_\gamma- p_{n+1}/\vep$ rather than on 
$p_\gamma + p_{n+1}$, Eq.~(\ref{cfact}) is not fulfilled by the definition
in Eq.~(\ref{fwrong}).

This effect can easily be understood.
In its collinear-fragmentation process the triggered parton (particle) turns out
to be necessarily less isolated (its energy decreases while the amount of
accompanying hadronic energy increases)
and, eventually, the isolation criterion can
be violated. Thus one cannot insist on factorizing a short-distance subprocess
that depends on the fixed isolation parameter $\varepsilon_h$. Hard partons
produced at short distances have to be more isolated than the triggered
particle $H$ and their isolation has to be increased as the energy 
of $H$ decreases. 

This is the key point to show that the isolated cross section 
is a jet-type observable. 
The isolation criterion in Eq.~(\ref{pic})
can indeed be recast in a form that explicitly fulfils collinear
factorizability by considering  $J = \{ \r, \d \}$ as the relevant parameters
of the hard-scattering subprocess. Here $\r$ is defined by
\beq{rc}
\r \equiv \frac{\x}{x_c} \geq \frac{2 \left( \sum\limits_{i\neq \gamma} E_i 
\;\Theta(\delta - \theta_{i\gamma}) + E_{\gamma}  \right)}{Q} > \x \;\;,
\eeq
and represents the upper limit on the total-energy fraction inside the
isolation cone. Using these isolation parameters, the measurement function 
that corresponds to the criterion in Eq.~(\ref{pic}) can be written as 
\beq{measfun}
F^{(n)}(Q,p_\gamma,J= \{ \r, \d \};p_1, \dots p_n) = 
\Theta\!\left( \r \frac{Q}{2} - 
\left[ \sum_{\stackrel{i=1}{i\neq \gamma}}^{n} E_i \; \Theta(\d -
\theta_{i\gamma}) + E_\gamma \right] \right) \;\;,
\eeq
Equation (\ref{cfact}), as well as Eqs.~(\ref{irsafe}) and (\ref{csafe}), 
straightforwardly apply to the function in Eq.~(\ref{measfun}) as long as
$\r > 2E_\gamma/Q=\x$. Note that, performing the limits of 
Eqs.~(\ref{irsafe})--(\ref{cfact})
the parameters $J= \{ \r, \d \}$ in Eq.~(\ref{measfun})
have to be kept {\em fixed} and regarded as 
variables that are independent of the momenta $p_\gamma,p_1, \dots p_n$. 
In particular, this implies that  
the factorized coefficient functions of Eq.~(\ref{fulltoJ})
explicitly depend on the parameters $J= \{ \r, \d \}$. 

We thus conclude that,
for the production cross section of isolated photons, factorization is valid
and  the {\em all-order} factorization formula is:
\beeq{facforiso'}
\!\!\!\!\!\!
\frac{1}{\sigma_0} 
\frac{d\sigma(Q^2, \x, x_c, \d)}{d\x} &=&
\sum_{a=q_f,{\bar q}_f,g} \; \int_{\x}^1 \frac{dx}{x} \;
\; D_{\gamma/a}(\x/x,\mu^2) \;
C_a(\as(\mu^2),Q^2/\mu^2;x, \r, \d) \nonumber \\
\label{facforiso}
&+&
\;C_\gamma(\as(\mu^2),Q^2/\mu^2;\x, \r, \d) \;\;,
\eeeq
where $C_a$ and $C_\gamma$ are the coefficient functions for the fragmentation
and direct components, respectively. Note that the fragmentation function
$D_{\gamma/a}$ is independent of the isolation parameters $\d$ and $\vep$.
The effect of the isolation is entirely included in $C_a$ and $C_\gamma$
through their dependence on $\d$ and $\r=\x/(1+\vep)$.

A final remark is in order. At present, a field-theory proof of the
factorization theorem is known only for fully inclusive
deep inelastic scattering (DIS). In the case of less inclusive observables,
factorization can be justified order by order in perturbation theory on the 
basis of general power-counting arguments \cite{CSS}.
The requirements of infrared and collinear safety and
collinear factorizability are sufficient to guarantee the validity
of factorization through power counting. In this respect, our
proof of factorization for isolated-photon cross sections cannot be considered
rigorous from a field-theory viewpoint, but it is certainly at the same level
of rigour as for any other semi-inclusive cross sections \cite{CSS}.

\subsection{Hadroproduction and photoproduction collisions}
\label{hadsect}

The discussion of the previous subsection can straightforwardly be extended to
the production of isolated photons in collisions of hadrons and/or real photons.
In these cases, the calculation of the cross section at the parton level 
contains additional singularities that are produced by initial-state collinear
radiation. These singularities have to be absorbed in the 
non-perturbative parton distributions of the colliding particles. Thus,
the essential difference with respect to $e^+e^-$ annihilation is that,
in hadronic collisions, the photon-isolation criterion must not spoil
the factorization of the initial-state collinear singularities.

As mentioned in Sect.~\ref{intro}, in the current experimental practice
\cite{exp} the hadronic-collision versions of the isolation criterion (\ref{pic})
involve transverse energies and angular distances evaluated in the 
azimuth--pseudorapidity plane. These variables are invariant under longitudinal
boosts along the beam direction and, hence, they are insensitive 
to initial-state collinear radiation. It follows that these types of isolation
criteria {\em fulfil} the QCD factorization theorem.

Note, however, that the factorization of initial-state collinear singularities
can be easily violated when boost-non-invariant variables are used. 
For instance, this is the case if one considers the angular distances
$R_{i\gamma}= \sqrt {\eta^2_{i\gamma} + \phi^2_{i\gamma}}$ and insist in using
energies rather than transverse energies.

\mysection{Fixed-order calculation in $e^+e^-$ annihilation}
\label{fixor}

To explicitly check the factorization formula (\ref{facforiso})
for $e^+e^-$ annihilation,
one needs a perturbative calculation to (at least) the first non-trivial 
order in $\as$. We have carried out this calculation in analytic form and
the results are presented in this section. We limit ourselves to 
the case in which the isolation parameters $\vep, \delta$ vary in the range
of practical interest, namely:
\beq{parr}
0 < \vep \leq 1 \;\;, \;\;\;\;\;\;  0 < \delta \leq \pi/2 \;\;.
\eeq

The coefficient functions $C_a$ for the fragmentation component of 
Eq.~(\ref{facforiso}) have
the following perturbative expansions:
\beq{perex}
C_a(\as(\mu^2),Q^2/\mu^2; x, \r, \d) = \lambda_a \left[
C_a^{(\LO)}(x, \r) \, + \, \asm \,C_a^{(\NLO)}(Q^2/\mu^2; x, \r, \d) \right]
+ {\cal O}(\as^2) \;\;,
\eeq
where $\lambda_{q_f}=\lambda_{{\bar q}_f} = e^2_{q_f}$,
$\lambda_{g}= \sum_f e^2_{q_f}$, $C_{q_f}=C_{{\bar q}_f}=C_q$
and $e_{q_f}$ is the electric charge of the quark of flavour $f$.

At the \LO\ only the quark coefficient function contributes:
\beq{loex}
C_q^{(\LO)}(x, \r) = \delta(1-x) \;\Theta(\r -1) \;, \;\;\;\;
C_g^{(\LO)}(x, \r) = 0 \;\;.
\eeq
Note also that $C_q^{(\LO)}(x, \r)$ is non-vanishing only for $\r \geq 1$,
i.e. for $\x \geq x_c$.

The results of our calculation for the
\NLO\ expressions can be written as 
\beeq{nlo}
C_a^{(\NLO)}(Q^2/\mu^2; x, \r, \d) &=&
\Theta(\r -1) \; \left [ \, C_a^{(\NLO , \,{\rm full})}(Q^2/\mu^2; x) 
\right. \nonumber \\
\label{nloex}
&+& \left. C_a^{(\NLO , \,{\rm in})}(x, \r, \d) \, \right] + \,
C_a^{(\NLO , \,{\rm out})}(x, \r, \d) \;\;,
\eeeq
where $C_a^{(\NLO , \,{\rm full})}(Q^2/\mu^2; x)$ denote the customary
\NLO\ coefficient functions \cite{aemp, nw} for the single-particle 
distribution, i.e. for the case in which no isolation is applied. Their
explicit expressions in the \msbar\ factorization scheme are:
\beeq{fullq'}
C_q^{(\NLO , \,{\rm full})}(Q^2/\mu^2; x) &=& C_F
\left\{ \left[ \frac{1+x^2}{1-x} \,\ln \frac{(1-x) Q^2}{\mu^2} 
- \frac{3}{2} \, \frac{1}{1-x} \right]_+ + \frac{1+x^2}{1-x} \,\ln x^2
\right. \nonumber \\
\label{fullq}
&~&\left. \;\;\; \;\;\; +
\left( \frac{2}{3} \pi^2 - \frac{11}{4} \right) \delta(1-x) 
+ \frac{5}{2} - \frac{3}{2} \; x
\right\} \;\;,
\eeeq
\beq{fullg}
C_g^{(\NLO , \,{\rm full})}(Q^2/\mu^2; x) = C_F \; 2 \; \frac{1+(1-x)^2}{x}
\, \ln \frac{(1-x) x^2 Q^2}{\mu^2} \;\;.
\eeq
The other two terms on the right-hand 
side of Eq.~(\ref{nloex}) derive from the isolation criterion and do not
depend on the factorization scheme. We find that:
\begin{itemize}
\item
The contributions $C_a^{(\NLO , \,{\rm in})}$ are non-vanishing only in the 
region
%\end{itemize}
\beq{psin}
%1 \leq 
\r \leq 1 + \tan^2 \frac{\d}{4} \;\;,
\eeq
and are given by
\beeq{qin}
\!\!\!\!&~& \!\!\!\!\!\!\!\! C_q^{(\NLO , \,{\rm in})}(x, \r, \d) = - \;
\Theta(x_\gamma^+ - x) \Theta(x - x_\gamma^-) \;C_F  \\
%\label{qin}
\!\!\!\!&\cdot& 
\left\{ \frac{1+x^2}{1-x} \,\ln \frac{(1-\r +x) (1-x) \tan^2 (\d/2)}{\r -1}
- (4-x) \left[ \frac{x - \r}{1-x}+ \frac{1}{1-x\sin^2(\d/2)}
\right] \right\} \;, \nonumber
\eeeq
\beeq{gin}
\!\!\!\!&& \!\!\!\!\!\!\!\!C_g^{(\NLO , \,{\rm in})}(x, \r, \d) = 
- \Theta(x_\gamma^+ - x) \Theta(x - x_\gamma^-) 
\;C_F  \\
%\label{qin}
\!\!\!\!&\cdot&\!\!\!\! 
\left\{ 2 \;\frac{1+(1-x)^2}{x} 
\,\ln \frac{(1-\r +x) (1-x) \tan^2 (\d/2)}{\r -1} 
%\right. \nonumber \\ && \left.  
- 4  \left[ x - \r + \frac{(1-x)}{1-x\sin^2(\d/2)}
\right] \right\} \;, \nonumber
\eeeq
where
\beq{xpm}
x_\gamma^{\pm} \equiv \frac{\r}{2} \left( 1 \pm  
\sqrt {1 - \frac{4(\r -1)}{\r^2 \sin^2(\d/2)}} \;\right) \;\;. 
\eeq
%\begin{itemize}

\item 
The contributions $C_a^{(\NLO , \,{\rm out})}$ 
are non-vanishing only in the region
%\end{itemize}
\beq{psout}
 \r < 1  \;\;,
\eeq
and have the following explicit expressions 
\beeq{qout}
\!\!\!&~& \!\!\!\!\!\!\!\! C_q^{(\NLO , \,{\rm out})}(x, \r, \d) = 
\Theta(\r - x) \;C_F  \\
\!\!\!&\cdot& 
\left\{ \frac{1+x^2}{1-x} \,\ln \frac{1}{(1-x) \tan^2 (\d/2)}
- \frac{4-x}{2} \left[ \frac{1}{1-x}- \frac{1+x \sin^2(\d/2)}{1-x\sin^2(\d/2)}
\right] \right\} \;, \nonumber
\eeeq
\beeq{gout}
\!\!\!\!&& \!\!\!\!\!\!\!\!C_g^{(\NLO , \,{\rm out})}(x, \r, \d) =
\Theta(\r - x) \;C_F  \\
\!\!\!\!&\cdot&\!\!\!\! 
\left\{ 2 \;\frac{1+(1-x)^2}{x} \,\ln \frac{1}{(1-x) \tan^2 (\d/2)}
%\right. \nonumber \\ && \left.  
- 2  \left[ 1 - \frac{(1-x)(1+x\sin^2(\d/2))}{1-x\sin^2(\d/2)}
\right] \right\} \;. \nonumber
\eeeq

\end{itemize}

The origin of the various \NLO\ contributions in
Eq.~(\ref{nlo}) is easily understood. We are interested in the process 
$\gamma^* \to q + \bar{q} + g + \gamma$ when one of the QCD partons
is collinear to the photon.
Owing to this collinear decay, to compute the coefficient function
$C_a^{(\NLO)}(Q^2/\mu^2; x, \r, \d)$, we simply have to evaluate the cross 
section\footnote{More technical details can be 
found in Ref.~\cite{AFGKP}.}
for the three-parton subprocess $\gamma^* \to q + \bar{q} + g$ when 
the triggered
parton $a$ ($a=q, \bar{q}$ or $g$) carries momentum $p$, parallel to
$p_\gamma$, and energy fraction $x= 2E/Q$. We
denote by $p_1,p_2$ the momenta of the other two partons and by $x_1,x_2$
their energy fractions. According to the isolation criterion specified
by Eq.~(\ref{measfun}), the corresponding measurement functions is:
\beq{measfun12}
F^{(2)}(Q,p,\{ \r, \d \};p_1,p_2) = 
\Theta\!\left( \r - \sum_{i=1,2} x_i \; \Theta(\d -
\theta_{i\gamma}) - x \right) \;.
\eeq
To make explicit the factorization of collinear singularities, we
rewrite Eq.~(\ref{measfun12}) by adding and subtracting a contribution
that is independent of the momenta, as follows:
\beq{measfun12sub}
\Theta\!\left( \r - 1\right) +
\left[ F^{(2)}(Q,p,\{ \r, \d \};p_1,p_2) - \Theta\!\left( \r - 1\right) \right]
\;\;.
\eeq 
When inserted into Eq.~(\ref{Jxs}) and combined with the virtual correction,
the first term in Eq.~(\ref{measfun12sub}) gives exactly 
(cf. Eq.~(\ref{fullxs})) the fully inclusive contribution
$C_a^{(\NLO , \,{\rm full})}(Q^2/\mu^2; x)$ to Eq.~(\ref{nlo}).

Then, we have to consider the term in the square bracket in 
Eq.~(\ref{measfun12sub}), which,
on the basis of the general factorization argument of
Sect.~\ref{factsec}, is expected to give a non-singular contribution.
To show that, we decompose 
this term in two parts
that correspond to the cases in which
one additional parton, either $p_1$ or $p_2$, is inside the isolation cone
$(F^{(2, {\rm in})})$ and both partons are outside it
$(F^{(2, {\rm out})})$. Using Eq.~(\ref{measfun12}), we obtain:
\beeq{measfun12inout'}
\left[ F^{(2)}(Q,p,\{ \r, \d \};p_1,p_2) - \Theta\!\left( \r - 1\right) \right]
&=& F^{(2, {\rm in})}(Q,p,\{ \r, \d \};p_1,p_2) \nonumber \\
\label{measfun12inout}
&+&
F^{(2, {\rm out})}(Q,p,\{ \r, \d \};p_1,p_2) \;,
\eeeq
where
\beq{measin}
F^{(2, {\rm in})}(Q,p,\{ \r, \d \};p_1,p_2) = 
- \left\{ \Theta(\d - \theta_{1\gamma}) \Bigl[ 
\Theta\!\left( \r - 1 \right)
- \Theta\!\left( \r - x_1 - x \right) \Bigr]
+ ( 1 \leftrightarrow 2 ) \right\} \;, \\
 \eeq
\beq{measout}
%\label{measout}
F^{(2, {\rm out})}(Q,p,\{ \r, \d \};p_1,p_2) =
\Theta(\theta_{1\gamma} - \d) 
\;\Theta(\theta_{2\gamma} - \d) \Bigl[ \Theta\!\left( \r -  x \right) -
\Theta\!\left( \r -  1 \right) \Bigr]
\;.
\eeq

Let us first consider the emission inside the cone.
Since $x_1 + x \geq 1$ because of kinematics, the term in the square bracket
on the
right-hand side of Eq.~(\ref{measin}) corresponds to the phase-space region
\beq{phspin}
x_1 + x > \r \geq 1 \;.
\eeq
This forbids the parton $p_1$ to become either soft ($x_1 = 0$) or 
collinear to the photon $(\theta_{1\gamma} =0)$
because in both cases the three-parton kinematics implies 
$x_1+x=1$, thus violating the constraint (\ref{phspin}). Therefore, we can
safely perform the integration over $p_1$ and obtain the finite contribution
$C_a^{(\NLO , \,{\rm in})}(x,\r,\d)$ in Eqs.~(\ref{nlo}), (\ref{qin}), 
(\ref{gin}). Note that parton
radiation inside the isolation cone is included in both
terms $C_a^{(\NLO , \,{\rm full})}$ and $C_a^{(\NLO , \,{\rm in})}$
of Eq.~(\ref{nlo}). The 
contribution of $C_a^{(\NLO , \,{\rm in})}$ is negative (cf.
Eq.~(\ref{measin}))
and represents the suppression effect of the non-isolated distribution produced
by the isolation criterion.

A similar discussion applies to the emission outside the cone.
Since $x \leq 1$ because of kinematics, the term in the square bracket on the 
right-hand side of Eq.~(\ref{measout}) vanishes when $\r \geq 1$.
Thus the term in the square bracket can be replaced by
\beq{phspout}
1 > \r \geq x \;.
\eeq
In this region, $p_1$ and $p_2$ cannot become either soft or
collinear because in both cases one has $x=1$. Therefore the
integration over $p_1$ and $p_2$ is safe and we obtain 
the finite contribution
$C_a^{(\NLO , \,{\rm out})}(x,\r,\d)$ in 
Eqs.~(\ref{nlo}), (\ref{qout}), (\ref{gout}).
Note that the $C_a^{(\NLO , \,{\rm out})}$ is positive,
although smaller (cf. the subtraction in Eq.~(\ref{measout}))
than the full non-isolated contribution 
$C_a^{(\NLO , \,{\rm full})}$ in the region $\r < 1$.

We remind the reader that the coefficient function for the direct component
of the factorization formula (\ref{facforiso}) was analytically computed
to the lowest order by Kunszt and Trocsanyi \cite{zoltan}.
Using our notation, their result can be written as follows:
\beq{cdirect}
C_\gamma(\as(\mu^2),Q^2/\mu^2; \x, \r, \d) = \frac{\alpha}{2\pi} \lambda_g
\left[ \frac{1}{C_F} \;C_g^{(\NLO)}(Q^2/\mu^2; \x, \r, \d) + 
{\cal O}(\as(\mu^2)) \right] \;,
\eeq 
where $\alpha$ is the fine structure constant and 
$C_g^{(\NLO)}(Q^2/\mu^2; \x, \r, \d)$ is given in Eq.~(\ref{nlo}). 
The relation (\ref{cdirect}) between $C_\gamma$ and our result for the
\NLO\ gluon coefficient function $C_g^{(\NLO)}$ can be regarded as a partial
check of the calculation described in this section.

Our \NLO\ results in Eqs.~(\ref{nlo})--(\ref{gout})
do not fully confirm those in Ref.~\cite{berger}. There, 
the coefficient functions $C_a^{(\NLO)}$ were 
found to be affected by singularities
(single and double poles in $1/\epsilon$, where $\epsilon=4-d$ parametrizes 
the number $d$ of
space-time dimensions in dimensional regularization) that would spoil
conventional factorization at the specific point $x_\gamma= x_c$.
The method of calculation used in Ref.~\cite{berger}
and the interpretation of these singularites
have been criticized in Ref.~\cite{AFGKP}.
The method described
in this section clearly exhibits the factorization of the collinear 
singularities. Using the decomposition in Eq.~(\ref{measfun12sub}),
we separate a term (that in the square bracket), which is manifestly free from
singularities\footnote{By `manifestly free from singularities' we mean the 
following. If this term is evaluated in $d=4-\epsilon$ space-time dimensions, 
then the limit $\epsilon \to 0$ can safely be performed because it leads
to the contributions 
$C_a^{(\NLO , \,{\rm in})}(x,\r,\d), C_a^{(\NLO , \,{\rm out})}(x,\r,\d)$
that are integrable in {\em any} interval of the kinematical variables 
$x,\r,\d$. This integrability guarantees that the perturbative coefficient
functions are well-defined distributions. However, it does not imply
that they are non-singular functions (cf. the discussion in 
Sect.~\ref{divsec}).}
and computable in 4 space-time dimensions,
from a remaining contribution whose dependence on the
isolation parameters is only due to the constraint $r_\gamma \geq 1$. This is
an overall constraint (i.e. it does not act on the partonic variables) and 
its effect is thus harmless: the ensuing (dimensionally regularized)
singularities are the collinear singularities that
universally enter the fully-inclusive cross section.

\setcounter{footnote}{0}
\mysection{Divergent behaviour for $x_\gamma \sim x_c$}
\label{divsec}

\begin{figure}
  \centerline{
    \setlength{\unitlength}{1cm}
    \begin{picture}(0, 8.5)
       \put(0,0){\includegraphics{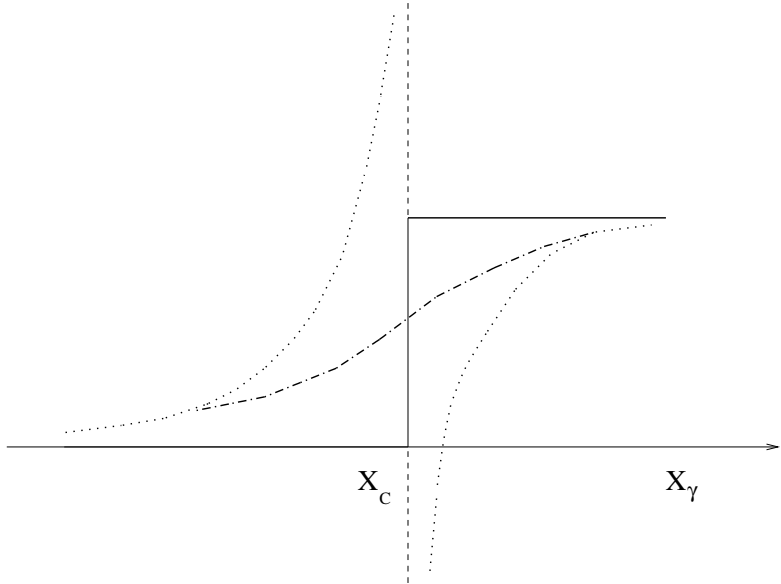}}
    \end{picture}}
  \caption[DATA]{The divergent behaviour of the isolated-photon cross section
in the vicinity of the critical point $x_\gamma=x_c$ at \LO\ (solid),
\NLO\ (dotted) and as expected after resummation (dot-dashed).  
}\label{fig_double}
\end{figure}

The \NLO\ coefficient functions $C_a^{(\NLO)}(Q^2/\mu^2; x, \r, \d)$ in
Eq.~(\ref{nlo}) are well-behaved for any value $\r \neq 1$, i.e. both for 
$\x < x_c$ and for $\x > x_c$. However, when $\r \to 1$ they become divergent.
The divergent behaviour (Fig.~\ref{fig_double})
at this point $\x = x_c$, which we shall call the
{\em critical point}, can be
easily derived from the explicit espressions in 
Eqs.~(\ref{qin}), (\ref{gin}) and (\ref{qout}), (\ref{gout}): 
\begin{itemize}
\item When the critical point is approached from below:
\beq{bel}
x_\gamma < x_c \;,
\eeq
we obtain
\beeq{qsingbel'}
C_q^{(\NLO)}(Q^2/\mu^2; x, \r, \d) &\!\!\!\!=\!\!\!\!& \delta(1-x) C_F
\left[ \ln^2 \left( (1-\r) \tan^2(\d/2) \right) \right. \nonumber \\
&\!\!\!\!+ \!\!\!\!& \left. \frac{3}{2}
\,\ln \left( (1-\r) \tan^2(\d/2) \right) \right] 
%\nonumber \\
\label{qsingbel}
%&\!\!\!\!+\!\!\!\!& \const \;, 
+ {\cal O}(1) \;,
\\
\label{gsingbel}
C_g^{(\NLO)}(Q^2/\mu^2; x, \r, \d) &\!\!\!\!=\!\!\!\!& {\cal O}(1) \;.
\eeeq
\item When the critical point is approached from above:
\beq{abo}
x_c \leq \x \;,
\eeq
we find
\beeq{qsinab'}
C_q^{(\NLO)}(Q^2/\mu^2; x, \r, \d) &=& C_F \left\{
\delta(1-x) \left[ - \ln^2 \left( \frac{\r-1}{\tan^2(\d/2)} \right) 
- \frac{3}{2}
\,\ln \left( \frac{\r-1}{\tan^2(\d/2)} \right) \right] \right.\nonumber \\
\label{qsingab}
&+& \left. \left( \frac{1+x^2}{1-x} \right)_+ 
\ln \left( \frac{\r-1}{\tan^2(\d/2)} 
\right) \right\} + {\cal O}(1) \;,\\
\label{gsingab}
C_g^{(\NLO)}(Q^2/\mu^2; x, \r, \d) &=& C_F \; 2  \frac{1+(1-x)^2}{x}
\; \ln \left( \frac{\r-1}{\tan^2(\d/2)} \right) 
+ {\cal O}(1) \;,
\eeeq
\end{itemize}
where ${\cal O}(1)$ stands for any term that is finite for $\r = 1$. Owing to
relation (\ref{cdirect}), the coefficient function 
$C_\gamma(\as(\mu^2),Q^2/\mu^2; \x, \r, \d)$ for the direct component diverges
when $x_c$ is approached from above.

The presence of these logarithmic
divergences was first pointed out in Ref.~\cite{berger}.
It is not in contradiction with the 
factorization of the collinear singularities \cite{AFGKP}.
Indeed, these two
types of singularities have different physical origins, as
discussed in rather general terms in Ref.~\cite{CW}
and recalled below.

The factorization theorem deals with infrared singularities that affect
the parton-level calculation for the {\em whole} range
(or for {\em finite} intervals) of the relevant kinematical 
variables, e.g. $\x, \delta$. These singularities are due to the long-distance
component of the scattering process that is intrinsically non-perturbative, in
the sense that it is {\em not suppressed} by some inverse power of $Q$ as the 
hard-scattering scale $Q$ increases.
The validity of the factorization theorem 
for the isolated-photon cross section guarantees that this non-perturbative
component can be described by the universal fragmentation function 
$D_{\gamma/a}(x,Q^2)$. The remaining contributions to the cross section
are short-distance-dominated. They consist of $i)$~power-suppressed terms (cf.
the term ${\cal O}((1/Q)^p)$ on the right-hand side of 
Eq.~(\ref{facforfull})) that are controlled by non-perturbative phenomena
and of $ii)$~a short-distance component, the coefficient functions 
$C_A$, which depends only logarithmically on $Q$ and is computable as a power
series expansion in $\as(Q^2)$.

However, the fact that the coefficient functions are perturbatively computable
does not imply that the coefficients of their perturbative expansion are
non-singular functions. In fact, in general these coefficients are singular
generalized functions or distributions that lead to finite quantities only when 
they are integrated with sufficiently smooth test functions. The
`plus'-distribution 
\beq{pdist}
w(x) = 2 C_F \left[ \frac{1}{1-x} \ln \frac{1}{1-x} \right]_+
\eeq
that enters in Eq.~(\ref{fullq}) 
is a well-known example of this type of singular generalized functions.
An analogous divergent behaviour, namely 
\beq{smalld}
C_a^{(\NLO)}(Q^2/\mu^2; x, \r, \d)
\simeq 
\Theta(1-\r) \Theta(\r-x) C_F \ln \frac{1}{\tan^2(\d/2)} \cdot
\cases { \frac{1+x^2}{1-x} \;\;\;\;\;\;\;\;\;\;\;\,(a=q) \;,\cr 
2 \,\frac{1+(1-x)^2}{x} \;\;\;\;(a=g) \;,\cr}
\eeq
is observed in the \NLO\ coefficient functions of Eq.~(\ref{nlo})
when the cone size $\delta \to 0$ at fixed $\r$. The double- and
single-logarithmic divergences in Eqs.~(\ref{qsingbel}), (\ref{gsingab}), 
(\ref{qsingab}) are integrable in any 
neighbourhood of the point 
$\x = x_c$ and essentially belong to the same type of singularities
\cite{Cmor}.
These singularities
are known as
divergences of the Sudakov type and,
in spite of the validity of the factorization theorem,
they are still produced by the
radiation of soft and/or collinear partons.

The main difference between the divergent behaviour in 
Eqs.~(\ref{fullq}) and (\ref{smalld})  
and that in Eqs.~(\ref{qsingbel}), (\ref{qsingab}), (\ref{gsingab}) is that
the former appears near the exclusive boundary of the phase space
(i.e. $x \to 1, \d \to 0$) while the latter occurs at a point $\x = x_c$
{\em inside the physical region} $0 < \x < 1$. The Sudakov singularities 
that arise at an exclusive boundary of the phase space are the most common
and extensively studied\footnote{See, for instance, 
Refs.~\cite{Cmor, Spro}
and references therein.}. They are due 
to the loss of balance between the virtual contributions and the radiative
tail of the real emission, which is strongly suppressed in this extreme
kinematic regime. Sudakov singularities inside the physical region
of the phase space have attracted less attention  
but are nonetheless quite common in jet physics. Some examples are the 
the $C$-parameter distribution \cite{ERT}
in $e^+e$ annihilation and the jet shape \cite{seymour}
in hadron collisions.

The general origin of Sudakov singularities at a critical point
inside the physical region has recently been discussed \cite{CW}.
They arise whenever the observable in question has a non-smooth behaviour
in some order of perturbation theory at that point. This can happen
if the phase-space boundary for a certain number of partons lies inside that
for a larger number, or if the observable itself is defined in a non-smooth
way. Both mechanisms are responsible for the singular behaviour 
in the case of the isolated-photon cross section.
The double- and single-logarithmic divergences in the square brackets of 
Eqs.~(\ref{qsingbel}), (\ref{qsingab}) are due to the first mechanism, 
while the remaining single-logarithmic terms in 
Eqs.~(\ref{qsingab}), (\ref{gsingab}) are due to the second mechanism. We 
discuss these points in turn.

At the \LO\ the 
phase-space region $\r < 1$ is not accessible to the fragmentation component
because there is a single parton,
the triggered quark, inside the isolation cone. Therefore the coefficient
function $C_q^{(LO)}(x,\r)$ in Eq.~(\ref{loex}) has a step at the critical point
$\r=1$. 
As proved in Ref.~\cite{CW},
this step-wise behaviour necessarily 
produces double-logarithmic singularities at the next order in perturbation 
theory. When applied to our case, the general argument of Ref.~\cite{CW}
is as follows. The \NLO\ subprocess
contains a term that is obtained by radiating a soft (and collinear) gluon from
the \LO\ subprocess. This term gives the following contribution to the
\NLO\ coefficient function
\beq{softem}
C_q^{(NLO)}(x,\r,\d) = \int_0^1 dz \;w(z) \;C_q^{(LO)}(x,\r(z)) + \dots \;\;, 
\eeq
where $1-z$ is the energy fraction of the soft gluon ($1-z \ll 1$) and
the dots stand for less singular terms when $\r \to 1$. In Eq.~(\ref{softem}), 
$w(z)$ denotes the
probability (in units of $\as/2\pi$) of emission of the soft gluon from the
triggered quark; it is given by the
usual `plus'-distribution in Eq.~(\ref{pdist}). 
The integration over $z$ 
follows from energy conservation and
relates the value of the parameter $\r$ after the emission (i.e. on the
left-hand side) to the corresponding value $\r(z)$ before the emission.
To explicitly evaluate Eq.~(\ref{softem}), we have to specify how $\r(z)$
depends on $\r$ and $z$.
Since $\r$ is
constrained by the fraction of the total energy inside the isolation cone (see
Eq.~(\ref{rc})), its value decreases $(\r < \r^{({\rm out})}(z))$
or increases $(\r > \r^{({\rm in})}(z))$
according to whether
the soft gluon is radiated outside or inside the cone. Neglecting less singular
terms\footnote{The actual size of the coefficient in front of the shift $(1-z)$ 
on the right-hand side of Eqs.~(\ref{rzout}), (\ref{rzin}) would affect only the
single-logarithmic contributions.} 
in the soft limit $1-z \ll 1$, we can thus write 
\beeq{rzout}
\r^{({\rm out})}(z) &\simeq& \r + (1-z) \;,\\
\label{rzin}
\r^{({\rm in})}(z) &\simeq& \r - (1-z) \;.
\eeeq
Using the explicit expression (\ref{loex}), (\ref{pdist}) 
and inserting Eqs.~(\ref{rzout}) and 
(\ref{rzin}) in the integral (\ref{softem}), we respectively obtain:
\beeq{singout'}
\!\!C_q^{(NLO, \,{\rm out})}(x,\r,\d) &\!\!=\!\!& 2 \;\delta(1-x) C_F
\int_0^1 dz \;\left(\frac{1}{1-z} \ln \frac{1}{1-z} \right)_+
\Theta(\r -z) \,+ \dots \nonumber \\
\label{singout}
&\!\!=\!\!& + \;\Theta(1-\r) \,\delta(1-x) \,C_F \ln^2 |\r-1| \,+ \dots \;\;, 
\eeeq
\beeq{singin'}
\!\!\!\!\!\!\!\!\!\!\!\!\!C_q^{(NLO, \,{\rm in})}(x,\r,\d) 
&\!\!\!=\!\!& 2 \;\delta(1-x) C_F
\int_0^1 dz \;\left(\frac{1}{1-z} \ln \frac{1}{1-z} \right)_+
\Theta(\r -1- (1-z)) \,+ \dots \nonumber \\
\label{singin}
&\!\!\!=\!\!& - \;\Theta(\r-1) \,\delta(1-x) \,C_F \ln^2 |\r-1| \,+ \dots \;\;,
\eeeq
in agreement with the double-logarithmic terms in Eqs.~(\ref{qsingbel}), 
(\ref{qsingab}).

Equations (\ref{softem}), (\ref{singout}), (\ref{singin}) explain the mechanism 
that produces the 
logarithmic divergences in the square brackets of 
Eqs.~(\ref{qsingbel}), (\ref{qsingab}).
The `plus'-prescription in Eq.~(\ref{pdist})
arises from adding {\em real} and {\em virtual} soft-gluon radiation
and, as a result of the cancellation of the soft singularities, leads to 
finite quantities whenever it acts on smooth functions of $z$.
However, this is not the case of
Eq.~(\ref{softem}), because $C_q^{(LO)}(x,\r(z))$ is discontinuous at 
$\r(z)=1$. The divergent behaviour at the critical point
is thus due to the imperfect compensation \cite{berger} between real and virtual
contributions, which occurs in the presence of the \LO\ step-like discontinuity
at $\x=x_c$.
Because of the different kinematic recoil in Eqs.~(\ref{rzout}) and 
(\ref{rzin}), at \NLO\ the cross section has a {\em double-sided}
singularity (Fig.~\ref{fig_double}), that is, it diverges to $+\infty$ and 
$-\infty$ below and above the critical point $\x=x_c$, respectively.

The single-logarithmic term\footnote{We recall that these single-logarithmic
divergences appear also in the \LO\ coefficient function $C_\gamma$ of the
direct component of the cross section.} in Eq.~(\ref{qsingab})
(outside the square bracket) 
and that in Eq.~(\ref{gsingab}) have a different origin.
They arise from the integration of the collinear spectrum of the parton that is
radiated by the triggered parton $a$ at an angle $\theta_{1\gamma}$ inside
the isolation cone. They have indeed the form:
\beq{slterm}
P_{qa}(x) \int_{\mu^2/Q^2}^{\theta_{\rm max}^2} 
\frac{d\theta_{1\gamma}^2}{\theta_{1\gamma}^2} = P_{qa}(x) 
\left( \ln \theta_{\rm max}^2 + \ln Q^2/\mu^2 \right) \;,
\eeq
where $P_{qa}(x)$ is the relevant Altarelli--Parisi probability
\beq{app}
P_{qq}(x) = C_F \left( \frac{1+x^2}{1-x} \right)_+ \;, \;\;\;\;
P_{qg}(x) = C_F \,\frac{1+(1-x)^2}{x} \;\;.
\eeq
The lower limit of integration over $\theta_{1\gamma}^2$ comes from the
factorization of the collinear singularity at $\theta_{1\gamma}=0$ in the
non-perturbative fragmentation function. The upper limit $\theta_{\rm max}^2$
comes from the kinematics of the process. Of course, $\theta_{1\gamma}$ has to
be smaller than the cone size $\d$. However, when $\r \to 1$ at fixed $x$ and
$\d$, the energy radiated inside the cone by the splitting process $q \to a$
can violate the isolation constraint before the cone boundary is actually
appproached by $\theta_{1\gamma}$. Therefore, $\theta_{\rm max}^2 \sim \r -1$
and Eq.~(\ref{slterm}) gives the single-logarithmic terms of 
Eqs.~(\ref{qsingab}) and (\ref{gsingab}). These terms are due to the non-smooth
character of the isolation criterion, which enforces sharp boundaries
on the energies and angles of the radiated partons.

\mysection{Outlook: higher orders, resummation  \newline
%\\
and non-perturbative effects}
\label{finsec}

In this paper we have shown that the factorization of collinear singularities 
in cross sections for the production of isolated particles defined by
the criterion (\ref{pic}) is valid to any order in QCD perturbation theory.
The non-perturbative component of the scattering process that is not 
power-suppressed at high transferred momentum $Q$ is thus taken into account
by the universal fragmentation functions $D_{H/a}(x,Q^{2})$, whereas the 
isolation condition is consistently included in the short-distance
subprocess.

In the case of isolated photons produced in $e^+e^-$ annihilation, 
we have checked
the factorization pattern by performing an explicit calculation at 
\NLO\ in $\as$. This calculation shows that, although infrared and collinear
safe, the short-distance component of the \NLO\ cross section 
has still a divergent behaviour (Fig.~\ref{fig_double})
when the photon energy fraction
$x_{\gamma}$ approaches a critical value $x_c=1/(1+\vep)$ that is  
located inside the physical region $0 < x_{\gamma} < 1$. As shown in
Eqs.~(\ref{qsingbel}), (\ref{gsingbel}), (\ref{qsingab}), (\ref{gsingab}),
the divergences are double-logarithmic. They are due to the loss of balance
between real and virtual contributions that is enforced by the
non-smooth character of the energy isolation criterion. In Sect.~\ref{divsec}
we have discussed in detail the physical mechanisms that produce these
singularities of the Sudakov type.
 
The same mechanisms leading to Eqs.~(\ref{qsingbel}), (\ref{gsingbel}),
(\ref{qsingab}), (\ref{gsingab})
will enhance the double-logarithmic divergences by further integer powers
of $\ln |\r - 1|$ in yet higher orders of perturbation theory. Since
these divergences are unphysical \cite{berger},
QCD calculations at {\em any} finite order in perturbation theory
cannot give reliable phenomenological predictions for the isolated-photon
cross section in the region around $\x = x_c$.

This problem may be overcome by trying to avoid the phase-space region
near the critical point. However, also in this case, some general theoretical
understanding of the phenomenon is necessary to assess the extent of the
dangerous region. The identification of the dangerous region in the
perturbative calculation is even more difficult in the case of 
hadron collisions \cite{berger}.
Here the singularities appear at a critical 
point in the partonic cross section and are smeared 
by the convolution 
with the parton distributions of the colliding hadrons.

A better way to deal with the problem is to identify and resum the soft-gluon
divergences to {\em all} orders in perturbation theory. The resummation
approach has been used successfully for the treatment of the Sudakov
singularities near the exclusive phase-space boundary for many observables
\cite{Cmor, Spro}.
The same approach was advocated in 
Refs.~\cite{berger, AFGKP}
for isolated photons and in Ref.~\cite{CW}
for a general treatment of the Sudakov singularities inside the physical 
region. 
 
\begin{figure}
  \centerline{
    \setlength{\unitlength}{1cm}
    \begin{picture}(0,7.5)
       \put(0,0){\includegraphics{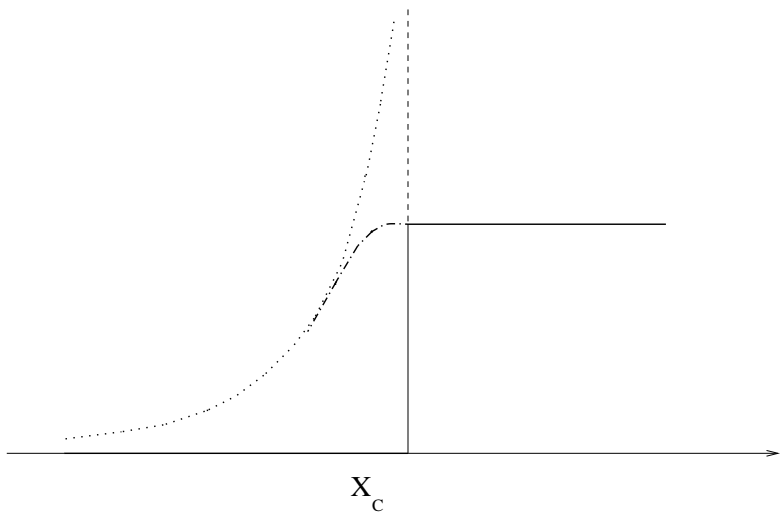}}
    \end{picture}}
  \caption[DATA]{Divergent behaviour at a critical point with single-sided
singularities: \LO\ (solid), \NLO\ (dotted). The dot-dashed line denotes
the Sudakov shoulder obtained after all-order resummation. 
}\label{fig_single}
\end{figure}

All-order resummation has already been carried out \cite{CW}
for a particular type of critical point inside the physical 
region.
This type of critical point (Fig.~\ref{fig_single})
regards cross sections that at \LO\ have a stepwise behaviour at that point
and at \NLO\
show a divergence on a {\em single} side of the step. After resummation, the
cross section is finite, continuous and differentiable at the critical point.
Rather than suppressing the \NLO\ divergence and thus mantaining the step-like
behaviour, 
resummation leads to the suppression of the step.
The general form of the resummed cross section is 
a smooth extrapolation from the region where the \NLO\ divergence appears
towards the critical point,
joining smoothly with the finite value that the cross section has on the other
side of the \LO\ step. This characteristic structure was called
a Sudakov shoulder \cite{CW}.

The resummation of the {\em double-sided} singularities 
(Fig.~\ref{fig_double}) of the isolated-photon
cross section is technically more complicated. We expect that it 
leads to a smooth `jump' structure 
obtained by smearing (convoluting) two Sudakov shoulders:
a shoulder below and an inverted shoulder above the critical point
$\x = x_c$. As a result, the two sides of the \LO\ step match
at some intermediate 
value at the critical point in the all-order distribution, and the step 
is smoothed. The actual derivation of the resummed calculation
requires more study,
since the type of smearing (convolution) to be applied to the two shoulders 
strongly depends on the detailed isolation kinematics.
Work on resummation is in progress, and the results will be reported
elsewhere.

Note that the perturbative divergences at $\x =x_c$ correspond to integrable 
singularities and therefore they could in principle be removed by 
non-perturbative smearing effects as expected, for instance, from
hadronization. However, 
since the hadronization smearing should cancel divergent terms
proportional to some power of $\as(Q)$,
this would require that the short-distance cross section contains
non-perturbative contributions that are not power-suppressed at large $Q$.
On the basis of our resummation argument, we do not
anticipate the presence of these contributions.
The resummation of soft-gluon effects
to all orders of perturbation theory should be sufficient to
render the isolated-photon cross section finite and smooth throughout the
physical phase space. This suggests that the non-perturbative contributions
that are not included in the fragmentation function  
are still power-suppressed.

\noindent {\bf Acknowledgements.} We are grateful to Patrick Aurenche and
Jean-Philippe Guillet for their collaboration at an early stage of this work.
We would like to thank Ed Berger for discussions and comments.

\end{document}